\documentstyle[12pt,psfig,epsf]{article}
\begin{document}
\thispagestyle{empty}
\hfill NTUA-2/01
\vspace{5mm}
\begin{center}

{\bf QED$_3$ WITH DYNAMICAL FERMIONS IN AN EXTERNAL MAGNETIC FIELD}

\vspace{0.5cm}

{\bf J. Alexandre$^a,$ K. Farakos$^a$, S.J. Hands$^b$, G. Koutsoumbas$^a$,\\ 
S.E. Morrison$^b$}
\bigskip

$^a$Department of Physics, National Technical University of
Athens,\\
Zografou Campus, 157 80 Athens, GREECE.
\bigskip

$^b$Department of Physics, University of Wales Swansea,\\ Singleton Park,
Swansea SA2 8PP, U.K.
\vspace{10mm}

{\bf Abstract}
\end{center}

In this paper, we present results of numerical lattice simulations of
two-flavor QED in three space-time dimensions. 
First, we provide evidence that chiral symmetry is spontaneously broken in the 
chiral and continuum limit. Next we discuss the role of an external 
magnetic field $B$ on
the dynamically generated fermion mass.
We investigate the $B$-dependence of the
condensate through calculations with dynamical fermions using the
non-compact formulation of the gauge field, and compare the 
results with those of
a comparable study using the quenched approximation. 

\vspace{5mm}

\newcommand{\be}{\begin{equation}}
\newcommand{\ee}{\end{equation}}
\newcommand{\bea}{\begin{eqnarray}}
\newcommand{\eea}{\end{eqnarray}}
\newcommand{\real}{{\rm l}\! {\rm R}}
\newcommand{\ra}{\rightarrow}
\newcommand{\tr}{\mbox{ tr }}
\newcommand{\Tr}{\mbox{ Tr }}
\newcommand{\al}{\alpha}
\newcommand{\bt}{\beta}
\newcommand{\bz}{{\bar{z}}}

\newcommand{\del}{\Delta}
\newcommand{\Th}{\Theta}
\newcommand{\td}{\tilde{\del}}
\newcommand{\g}{\gamma}
\newcommand{\m}{\mu}
\newcommand{\n}{\nu}
\newcommand{\bchi}{\bbox{\chi}}
\newcommand{\nn}{\nonumber}

\section{Introduction}
\label{sec:intro}

The intriguing phenomenon of high $T_c$ superconductivity has 
inspired several models, consisting mostly of elaborations
of a doped version of the Hubbard model.
The model presented in \cite{NN} claims the generation 
of superconductivity 
through a ``mixed Chern-Simons term" without the need to introduce a 
local gauge symmetry breaking condensate. The model invokes the idea of
electron substructure (spin-charge separation) considering the low-energy
excitations of the system to be bosonic degrees of freedom called spinons 
and fermionic ones called holons. The role of the phonon interactions crucial
to the formation of Cooper Pairs in BCS superconductors is now
played by a ``statistical" gauge field, which arises from the underlying
dynamics of the holons hopping between lattice sites. More specifically, the
model is based on a so-called $\tau_3$-U(1) interaction of the 
fermions with the gauge field. 

An alternative form of the model has been proposed in \cite{fm}.
The model is meant to represent the
na\"\i ve continuum limit of the
effective low-energy Hamiltonian of the large-$U$, doped Hubbard model.
The action of the model reads:
\be
{\cal L} = -\frac{1}{4}(F_{\mu\nu})^2
-\frac{1}{4}\mbox{tr}({\cal G}_{\mu\nu})^2  +{\bar \Psi}D_\mu\gamma_\mu\Psi
\label{su2action}
\ee
where $D_\mu = \partial_\mu -ig_1A_\mu^S-ig_2\tau^aW_{a,\mu}
-i\frac{e}{c}A_\mu^{ext}$, 
with $A_\mu^S$ the statistical U(1)$_S$ gauge field, 
$W_\mu^a$ is the gauge potential of a local SU(2) group,
and ${\cal G}^a_{\mu\nu}$ 
the corresponding field strength. The $\Psi$ fields are
two-component spinors. 
The difference from the previous model lies in the fact that now the
full gauge symmetry 
is $\mbox{U}(1)_S\otimes \mbox{SU}(2),$ where we have denoted by 
U(1)$_S$ the abelian part of the symmetry. We call it a ``statistical" 
gauge field since from the physical point of view it describes 
statistical properties of the model, such as the doping, and not 
real electromagnetism. 

The strongly-coupled U(1)$_S$ interaction promotes the
spontaneous generation of a fermion -- anti-fermion condensate resulting in a
mass gap.
Arguments based on energetics
prohibit the generation of a parity-violating
but SU(2) gauge invariant term~\cite{vafa}, in favour of
a parity-conserving condensate, which necessarily breaks 
the SU(2) group down to a $\tau_3$-U(1) sector \cite{NN}.
The mass gap thus
breaks the SU(2) part down to an abelian sub-group which is the
analog of the $\tau_3$-U(1) in the model of \cite{NN}. 
The symmetry breaking pattern here follows the scheme:
$\mbox{U(1)}_S\otimes\mbox{SU}(2) \rightarrow \mbox{U(1)}_S \otimes 
\mbox{U}(1)_{\tau_3}.$

In equation (\ref{su2action}) we have also allowed for an additional 
U(1)$_{em}$ 
coupling, which describes the coupling to an external electromagnetic field
$A_\mu^{ext}$.
Actual {\it superconductivity} is obtained upon coupling
the system to such external electromagnetic potentials.
We now discuss the superconducting consequences of
the above dynamical breaking patterns of the SU(2) group.
Upon the opening of a mass gap in the fermion (hole) spectrum, 
one obtains a non-trivial result for the following Feynman matrix element:
${\cal S}^a = \langle W^a_\mu|J_\nu|0\rangle,~a=1,2,3$, with $J_\mu ={\bar
\Psi}\gamma _\mu \Psi $ the fermion-number current.
Due to the colour-group structure, only the massless $W^3_\mu $
gauge boson of the SU(2) group, corresponding to the $\tau _3$
generator in two-component notation, contributes to the
matrix element. The non-trivial result for the matrix element 
${\cal S}^3$ arises from an {\it anomalous one-loop graph}, 
and is given by \cite{RK,NN}:
\be
    {\cal S}^3 = \langle W^3_\mu|J_\nu|0\rangle=
({\rm sgn}{M})\epsilon_{\mu\nu\rho}
\frac{p_\rho}{\sqrt{p_0}}
\label{matrix2}
\ee
where $M$ is the parity-conserving fermion mass
generated dynamically
by the U(1)$_S$ group. As with 
other Adler-Bell-Jackiw anomalous graphs in gauge theories,
the one-loop result (\ref{matrix2})
is {\it exact} and receives no contributions from higher loops \cite{RK}.
This unconventional
symmetry breaking (\ref{matrix2}),
does {\it not have a local order parameter} \cite{RK,NN}
since its formation is inhibited by strong phase fluctuations,
thereby resembling  the
Kosterlitz-Thouless
mode of symmetry breaking. The {\it massless} gauge boson
$W_\mu^3$ of the
unbroken $\tau_3$-U(1) subgroup of SU(2) is responsible for the
appearance of a massless pole in the electric current-current
correlator, which is the characteristic feature
of any superconducting theory. As discussed in ref. \cite{NN},
all the standard properties of a superconductor, such as
the Meissner
effect, infinite conductivity, flux quantization, the London limit of
the action etc. are
recovered in such a case.
The field $W^3_\mu$, or rather its {\it dual} $\phi$ defined by
$\partial _\mu \phi \equiv \epsilon_{\mu\nu\rho}\partial_\nu W^3_\rho$,
can be identified with the Goldstone
boson of the broken U(1)$_{em}$ electromagnetic symmetry \cite{NN}.

If an external magnetic field is coupled to the charged excitations
about the superconducting state then new phenomena appear. 
In refs. \cite{fm,lattice} it is argued that an external magnetic field 
in this (2+1)-dimensional system induces the opening
of a second superconducting gap at the nodes of the $d$-wave gap, in
agreement with recent experimental results on the behaviour of the thermal
conductivity of high-temperature cuprates under the influence of strong
external magnetic fields \cite{krish}.

In this work we will study the phenomenon of the enhancement of the 
fermion condensate in the presence of an external magnetic field.
In this respect it is not important to gauge the SU(2) group. 
We study QED$_3$, a model which in the continuum has a local U(1) symmetry and
a global SU(2) symmetry. 
However, the lattice discretization we employ in order to perform
non-perurbative calculations restricts the global symmetry to a U(1) ``chiral''
symmetry,
which in turn is totally broken by the fermion -- anti-fermion condensate
$\langle\bar\Psi\Psi\rangle$.
The question of whether this symmetry is spontaneously broken in
QED$_3$, ie. 
whether $\langle\bar\Psi\Psi\rangle\not=0$ in the chiral limit bare fermion 
mass $m\to0$ and the
continuum (weak coupling) limit $g\to0$, 
has been an important issue in non-perturbative field
theory for many years. Initial studies based on Schwinger-Dyson (SD)
studies using the
photon propagator derived from the leading order $1/N_f$ expansion, where $N_f$
is the number of fermion flavors, suggested 
that for $N_f$ less than some critical value $N_{fc}$ 
the answer is positive 
\cite{appelquist}, with $N_{fc}\simeq3.2$. The model in the limit
$N_f\to N_{fc}$ is supposed to undergo an infinite-order phase transition
\cite{Mir}. Other studies 
taking non-trivial vertex corrections into account predicted chiral symmetry 
breaking for arbitrary $N_f$ \cite{pennington}. More recent studies which treat 
the vertex consistently in both numerator and denominator of the SD equations
have found $N_{fc}<\infty$, with a value either in agreement with the 
original study
\cite{maris}, or slightly higher $N_{fc}\simeq4.3$ \cite{nash}.
Finally a recent argument based on a thermodynamic inequality has predicted
$N_{fc}\leq{3\over2}$ \cite{ACS}. 

There have also been numerical attempts to resolve the issue via simulations of
non-compact lattice QED$_3$. Once again, opinions have divided 
on whether $N_{fc}$ is
finite and $\simeq3$ \cite{kogut}, or whether chiral symmetry is broken for all
$N_f$ \cite{azcoiti}. The principal obstruction to a definitive answer has been
large finite volume effects resulting from the presence of a massless photon in
the spectrum, which prevent a reliable extrapolation to the thermodynamic limit.
In this context it is also worth mentioning the three-dimensional 
Thirring model, which has a current-current
interaction mediated in the UV limit by a propagator of 
identical form in the large-$N_f$
limit to that of QED$_3$; moreover chiral symmetry breaking
for small $N_f$ is predicted by the same set of SD equations as those of QED$_3$
\cite{itoh}. This time, however, 
the critical behaviour is governed by a UV-stable fixed point rather than an IR
one, and 
numerical studies are more controlable, yielding values of $N_{fc}\simeq5$
\cite{thirring}.

The rest of the paper is organised as follows. In section \ref{sec:symm} we
outline the formulation of continuum QED$_3$ in Euclidean metric, and identify 
its global symmetries. In section \ref{sec:latt} we do the same for the lattice
version of the model. We also review the formulation of a homogeneous
background magnetic field $B$ in abelian lattice gauge theory.
Our numerical results are presented in section \ref{sec:results}.
First for $B=0$, we attempt an 
extrapolation to both continuum and chiral limits, and observe evidence for 
a non-vanishing
(though unexpectedly small) dimensionless chiral condensate, 
suggesting $N_{fc}>2$.
We are also able to estimate the combination of lattice sizes and bare masses
required for quantitative accuracy in further work. This work is described in 
section~\ref{subs:B=0}. In section~\ref{b.neq.0} we turn to 
non-zero magnetic field $B$, and study the resulting
enhancement of the condensate. This is the first lattice simulation of
QED$_3$ with $B\not=0$ using dynamical fermions; the problem has previously been
studied in the quenched approximation \cite{fkmm}. In particular we compare the behaviours of 
quenched versus dynamical fermions and calculate the condensate as a function 
of the magnetic field, and find different behaviours both in the strong
and the weak field regimes, and in strong and weak U(1)$_S$ coupling regimes.
We conclude with a study of the 
dependence of the condensate in the strength of the statistical gauge
interaction in the chiral limit, and uncover an important finite volume effect.

\section{The model in the continuum and its symmetries}
\label{sec:symm}

The three-dimensional 
continuum Lagrangian describing QED$_3$ is given (in Euclidean
metric, which we use hereafter) by ~\cite{farak,fm}, 
\be 
{\cal L} = \frac{1}{4}F_{\mu\nu}^2 
+ {\bar \Psi}_i D_\mu\gamma _\mu \Psi_i +m {\bar
 \Psi}_i \Psi_i 
\label{contmodel}
\ee
where $D_\mu = \partial _\mu -ig A^S_\mu - i e A^{ext}_\mu$,
and $F_{\mu\nu}$, ${\cal G}_{\mu\nu}$ are the corresponding field
strengths for an abelian U(1)$_S$ statistical gauge field $A^S_\mu$ and 
a background U(1)$_{em}$ gauge field $A^{ext}_\mu$, respectively. 
The fermions $\Psi_i$, $i=1,\ldots,N_f$,
are now four-component spinors. 
We note that $\Psi_i$ may be written in terms of the two-component spinors 
used in section \ref{sec:intro} as 
\be
\Psi_i \equiv \left( \begin{array}{c} \Psi_{i1} \\ 
\Psi_{i2}\end{array}\right).
\ee
A convenient representation for 
the $\gamma _\mu $, $\mu =0,1,2$, matrices 
is the reducible $4\times4$ representation of the 
Dirac algebra in three dimensions~\cite{appelquist} (in Euclidean 
metric we choose hermitian Dirac matrices):
\be
\gamma^0=\left(\matrix{\sigma_3&\phantom{-}{\bf0}\cr
{\bf0}&-\sigma_3\cr}\right);
\qquad\gamma^1=\left(\matrix{\sigma_1&\phantom{-}{\bf0}\cr
{\bf0}&-\sigma_1\cr}\right); 
\qquad\gamma^2=\left(\matrix{\sigma_2&\phantom{-}{\bf0}\cr
{\bf0}&-\sigma_2\cr}\right),
\label{reduciblerep}
\ee
where 
$\sigma_i$ are $2 \times 2$ Pauli matrices.
Then the Lagrangian (\ref{contmodel}) decomposes into two parts,
one for $\Psi_{i1}$ and one for $\Psi_{i2},$ 
which will be called ``fermion species" in the sequel.
The presence of an even
number of fermion species allows us to define chiral symmetry and
parity in three dimensions \cite{appelquist}, which we discuss below.
The bare mass $m$ term is parity conserving and  
has been added by hand 
in the Lagrangian (\ref{contmodel}). 
This term is generated dynamically anyway via 
the formation of the fermion condensate
$\langle{\bar \Psi} \Psi \rangle$
by the strong 
U(1)$_S$ coupling. Here we include such a term explicitly, since 
it is necessary from the technical point of view.

As is well known~\cite{appelquist} there exist two $4 \times 4 $ matrices 
which anticommute with $\gamma _\mu$, $\mu=0,1,2$: 
\be
\gamma_3=\left(\matrix{{\bf0}&{\bf1}\cr{\bf1}&{\bf0}\cr}\right),\qquad
\gamma_5=i\left(\matrix{\phantom{-}{\bf0}&{\bf1}\cr-{\bf1}&{\bf0}\cr}\right),
\label{gammamatr}
\ee
where the substructures are $2 \times 2$ matrices.
These are the generators of the `chiral' symmetry for 
the massless fermion theory: 
\bea
&~&   \Psi \rightarrow \exp(i\theta \gamma _3) \Psi, \qquad
\bar\Psi\rightarrow\bar\Psi\exp(i\theta\gamma_3);\nn \\
&~&   \Psi \rightarrow \exp(i\omega \gamma _5) \Psi,\qquad 
\bar\Psi\rightarrow\bar\Psi\exp(i\omega\gamma_5).
\label{chiral}
\eea
Note that these transformations do not exist in the 
two-component representation
of the three-dimensional Dirac algebra, and therefore 
the above symmetry is valid for theories 
{\em with even numbers of fermion species only.}

For later use we note that
the Dirac algebra in $(2 +1)$ dimensions satisfies the identities:
\bea 
&~&\gamma ^\mu \gamma ^\nu = -\delta ^{\mu\nu} 
- \tau_3 \epsilon ^{\mu\nu\lambda} \gamma ^\lambda \quad ; \quad  
\tau_3\equiv i\gamma_3\gamma_5=\left(\matrix{{\bf1}&\phantom{-}{\bf0}\cr
{\bf0}&-{\bf1}}\right)\nn\\
&~& \gamma ^\mu \gamma ^\lambda \gamma ^\mu = \gamma ^\lambda \nn \\
&~&\gamma ^\mu \gamma ^0 \gamma ^\rho \gamma ^\sigma \gamma ^\mu 
= - \delta ^{\rho \sigma}\gamma ^0 - 3 \tau_3 \epsilon ^{\rho \sigma} \nn \\
&~& \gamma ^\mu \gamma ^\rho \gamma ^\sigma \gamma ^\mu 
= - 3\delta ^{\rho \sigma} - \tau_3 \gamma ^0 \epsilon ^{\rho \sigma} \nn \\ 
&~&\gamma ^\mu \gamma ^\sigma \gamma ^\rho \gamma ^\tau \gamma ^\mu 
= - \delta ^{\rho \sigma}\gamma ^\tau 
- \delta ^{\rho \tau}\gamma ^\sigma + \delta ^{\sigma \tau}\gamma ^\rho
\label{identities}
\eea
which are specific to three dimensions. Here the Greek letters 
are spacetime indices, and repetition implies summation. 

{\it Parity} in this formalism is defined as the transformation:
\bea
P:~\Psi (x^0, x^1, x^2) &\rightarrow&-i\gamma_3\gamma_1 \Psi (x^0, -x^1, x^2); 
\nn\\\
\bar\Psi(x^0,x^1,x^2)&\rightarrow&\bar\Psi(x^0,-x^1,x^2)
i\gamma_1\gamma_3,
\label{parity}
\eea 
and it is easy to see that the parity-invariant mass term
$\bar\Psi\Psi$ 
corresponds to the species having masses with {\it opposite} signs 
~\cite{appelquist}, whilst a parity-violating one $\bar\Psi\tau_3\Psi$
corresponds to masses of the same sign. 

The set of generators 
\be
{\cal G} = \{ {\bf 1}, \gamma _3, \gamma _5, 
\tau_3\}
\label{generators}
\ee
form~\cite{farak,fm} 
a global U(2) $\simeq$ SU(2)$\otimes$U(1) symmetry. 
To see this define $\tilde\Psi\equiv\bar\Psi\tau_3$.
The U(2) symmetry is then
\be
\Psi\rightarrow U\Psi\qquad\tilde\Psi\rightarrow\tilde\Psi U^\dagger\qquad
U\in\mbox{U(2)}.
\ee
The identity matrix {\bf 1} generates the U(1) subgroup, 
while 
the other three form the SU(2) part of the group. 
The currents corresponding to the SU(2) transformations 
are:
\be
   J_\mu^\Gamma = {\bar \Psi} \gamma _\mu \Gamma \Psi 
\qquad \Gamma =\gamma _3,\gamma _5, \tau_3
\label{currents}
\ee
and are conserved in the absence of a fermionic mass
term. 
It can be readily verified that the corresponding charges
$Q_\Gamma \equiv \int d^2x \bar\Psi\gamma_0\Gamma \Psi $ lead
to an SU(2) algebra~\cite{farak}:
\be
[Q_3, Q_5]=2iQ_{\tau_3}, \qquad [Q_5,Q_{\tau_3} ]=2iQ_3, \qquad
[Q_{\tau_3}, Q_3]=2iQ_5.
\label{chargealgebra}
\ee
In the presence of a mass term, these currents are not all conserved:
\be
       \partial ^\mu J_\mu^\Gamma = 2m {\bar \Psi } \Gamma \Psi,
\label{anomaly}
\ee
for $\Gamma=\{\gamma_3,\gamma_5\}$,
while the currents corresponding to $\{{\bf 1},\tau_3\}$ 
are {\it always } conserved, even in the presence of a fermion 
mass. The situation parallels the treatment of the SU(2)$\otimes$SU(2)
chiral symmetry breaking in low-energy QCD and the partial conservation of
axial current (PCAC).
The bilinears
\bea
&~&{\cal A}_1 \equiv {\bar \Psi}\gamma _3 \Psi,  
\qquad {\cal A}_2 \equiv {\bar \Psi}\gamma _5 \Psi,  
\qquad {\cal A}_3 \equiv {\bar \Psi}\Psi 
\nn \\
&~&B_{1\mu} \equiv {\bar \Psi}\gamma _\mu \gamma _3 \Psi,~
B_{2\mu} \equiv {\bar \Psi}\gamma _\mu \gamma _5 \Psi,~
B_{3\mu} \equiv {\bar \Psi}\gamma _\mu \tau_3 \Psi,~\mu=0,1,2      
\label{triplets}
\eea
transform as {\it triplets} under SU(2). 
The SU(2) singlets are 
\be 
{\cal A}_4 \equiv {\bar \Psi}\tau_3 \Psi, \qquad 
B_{4,\mu} \equiv {\bar \Psi}\gamma _\mu \Psi 
\label{singlets}
\ee
i.e. the singlets are the parity-violating mass term, 
and the four-component fermion current. 

We now notice that in the case 
where the fermion condensate ${\cal A}_3$ is generated 
dynamically, energetics 
prohibit the generation of a parity-violating 
SU(2)-invariant term~\cite{vafa}, and so 
a parity-conserving mass term necessarily breaks~\cite{fm}
the $U_S(1)\otimes$SU(2) group down to a $U_S(1)\otimes$U(1)$_{\tau_3}$ 
group~\cite{NN}.

Finally, we note that were the global symmetries described here to be promoted
to local ones, as they are in both the $\tau_3$-U(1) model of \cite{NN} and the
$U_S(1)\otimes$SU(2) model of \cite{fm}, then the measure of the Euclidean
functional integral (i.e. the fermion determinant) would not be positive
definite. Positivity of the theory defined by (\ref{contmodel}) follows because
there exist matrices $\gamma_3$, $\gamma_5$ which anticommute with
$D_\mu\gamma_\mu$; hence the imaginary eigenvalues of this antihermitian
operator necessarily
occur in conjugate pairs. This property no longer holds once, e.g., the 
U(1)$_{\tau_3}$ symmetry is gauged. Thus this class of
models suffer from a `sign problem' in Euclidean metric, 
a feature shared with other potentially superconducting systems such as QCD at
high baryon number density \cite{HM}.

\section{Lattice formulation}
\label{sec:latt}

We now proceed with a description of the lattice formulation of the 
problem. The lattice action using staggered lattice fermion fields 
$\chi,\bar\chi$
is given by the formul\ae\/ below:
\bea
&~&S =\frac{\beta_G}{2} \sum_{x,\mu<\nu} F_{\mu \nu}(x) F^{\mu \nu}(x) 
+ \sum_{x,x^\prime} {\bar \chi}(x) Q(x,x^\prime) \chi(x^\prime)
\label{eq:action}
\eea
$$
F_{\mu \nu}(x) \equiv \alpha^S_\mu(x)+\alpha^S_\nu(x+\hat \mu)
-\alpha^S_\mu(x+\hat \nu)-\alpha^S_\nu(x)
$$
$$
Q(x,x^\prime) \equiv  m \delta_{x,x^\prime}+\frac{1}{2} \sum_\mu\eta_{\mu}(x) 
[\delta_{x^\prime,x+\hat \mu} U_{x\mu} V_{x\mu} 
-\delta_{x^\prime,x-\hat \mu} U_{x-\hat \mu,\mu}^\dagger 
V_{x-\hat \mu,\mu}^\dagger].
$$
The indices $x,~x^\prime$ consist of three integers 
$(x_1,~x_2,~x_3)$
labelling the lattice sites, where the timelike direction is now considered
as the third. Since the gauge action $F^2$ is unbounded from above, 
(\ref{eq:action})
defines the {\sl non-compact\/} formulation of lattice QED. 
The $\eta_\mu(x)$ are Kawamoto-Smit phases $(-1)^{x_1+\cdots+x_{\mu-1}}$
designed to ensure relativistic covariance in the continuum limit. 
For the fermion fields antiperiodic boundary conditions are used in 
the timelike direction and periodic 
boundary conditions in the spatial directions.
The phase factors in the fermion bilinear are defined by
$U_{x\mu} \equiv
\exp(i\alpha^S_{x \mu}),~~V_{x \mu} 
\equiv \exp(i \alpha^{ext}_{x \mu})$, where
$\alpha^S_{x \mu}$ represents the statistical gauge
potential and $\alpha^{ext}_{x \mu}$ the
external electromagnetic potential. In terms of continuum quantities,
$\alpha_{x\mu}^S= a gA_\mu^S(x)$, the coupling
$\beta_G \equiv \frac{1}{g^2 a}$, and
$\alpha_{x\mu}^{ext}=aeA_\mu^{ext}(x)$,
where $a$ is the physical lattice spacing.
The coupling $e$ to the external electromagnetic field is 
dimensionless.

The numerical results in this paper were obtained by simulating the 
action (\ref{eq:action}) using a standard Hybrid Monte Carlo (HMC) algorithm.
The form of (\ref{eq:action}) permits an even-odd partitioning, 
which means that a single flavor of staggered
fermion can be simulated. 
In the long wavelength limit, this can be shown to correspond to $N_f=2$
flavors of the four-component fermions $\Psi$ considered in 
section \ref{sec:symm}
\cite{BB}. Chiral condensation $\langle\bar\Psi\Psi\rangle
\not=0$ would then result in a pattern
$\mbox{U}(2N_f)\rightarrow\mbox{U}(N_f)\otimes\mbox{U}(N_f)_{\tau_3}$. 
For non-zero
lattice spacing $a$, however, the global symmetries for $m=0$ 
are only partially 
realised \cite{thirring}. In this case the symmetry is
\bea
\bar\chi_o\rightarrow\bar\chi_o e^{i\alpha}&~&\qquad
\chi_e\rightarrow e^{-i\alpha}\chi_e \nn\\
\bar\chi_e\rightarrow\bar\chi_e e^{i\beta}&~&\qquad
\chi_o\rightarrow e^{-i\beta}\chi_o,
\label{eq:latsym}
\eea
where $\chi_{o/e}$ denotes the field on odd (i.e. $
\varepsilon(x)\equiv(-1)^{x_1+x_2+x_3}=-1$)
and even sublattices respectively. Chiral symmetry breaking characterised by
$\langle\bar\chi\chi\rangle\not=0$ now has the pattern
$\mbox{U}(N)\otimes\mbox{U}(N)\rightarrow\mbox{U}(N)$
for $N$ flavors of staggered fermion.

We should now construct a lattice
version of the homogeneous magnetic field. This has already been 
done before in \cite{damhel} in connection with the abelian Higgs model.
We follow a slightly different prescription,
which we describe below \cite{lattice}, and which is closely related to 
a prescription for introducing a uniform topological charge density
in two dimensions \cite{smitvink}.

Since we would like to impose an external homogeneous
magnetic field ${\bf B=\nabla\times\alpha}^{ext}$
in the (missing) $z$ direction, we
choose the external gauge potential in such a way that the plaquettes in
the $x_1 x_2$ plane equal $B$, 
while all other plaquettes equal zero.
One way in which this can be achieved is through the choice:
$\alpha^{ext}_3(x_1, x_2, x_3)=0, {\rm for~all }~x_1,~x_2,~x_3,$ and
\bea
&~&\alpha^{ext}_1(x_1, x_2, x_3) =
\cases{\phantom{(N+1}-\frac{B}{2} (x_2-1),&$x_1 \ne N$;\cr  
      -\frac{B}{2}(N+1)(x_2-1),&$x_1=N$:\cr}
\nn\\
&~&\alpha^{ext}_2(x_1, x_2, x_3) =
\cases{\phantom{(N+2}+\frac{B}{2} (x_1-1),&$x_2 \ne N$;\cr  
       +\frac{B}{2}(N+1)(x_1-1),&$x_2=N$.\cr}
\eea
The lattice extent is assumed to be $N\times N$ in the $x_1x_2$ plane.
It is trivial to check that all plaquettes starting at $(x_1,x_2,x_3),$
with the exception of the one starting at $(N,N,x_3),$ equal $B$. The latter 
plaquette equals $(1-N^2) B = B -(N^2 B).$ One may say that the flux
is homogeneous over the entire $x_1x_2$ cross section of the lattice and
equals $B.$ The additional flux of $-(N^2 B)$ can be understood by the
fact that the lattice is a torus, that is a closed surface, and the 
Maxwell equation
${\bf \nabla \cdot B} =0$ implies that the total magnetic flux 
through the $x_1x_2$ plane
should vanish. This means that, if periodic boundary conditions are
used for the gauge field, the total flux of any configuration should be 
zero, so the (positive, say) flux $B,$ penetrating
the majority of the plaquettes will be accompanied by a compensating
negative flux $-(N^2 B)$ located on a single plaquette.
This compensating flux should be ``invisible", that is it should have no
observable physical effects. This is the case if the flux is an integer
multiple of $2 \pi: N^2 B =n 2 \pi \Rightarrow B=n \frac{2 \pi}{N^2},$ 
where $n$ is an integer. Thus we may say (disregarding the invisible 
flux) that the magnetic field is homogeneous over the entire cross section
of the lattice.\footnote{To check this translational invariance we 
measured the fermion condensate at every point in the $x_1 x_2$ plane. 
The results were the same at all points within the error bars, 
confirming homogeneity.} The integer $n$ may be chosen to lie in the 
interval $[0, \frac{N^2}{2}],$ 
since the model with integers between $\frac{N^2}{2}$ 
and $N^2$ is equivalent to that with integers 
taking on the values $N^2-n$.
It follows that the magnetic field 
strength $B$ in lattice units lies between 0 and $\pi.$ 

\begin{figure}[t]
\centerline{\psfig{figure=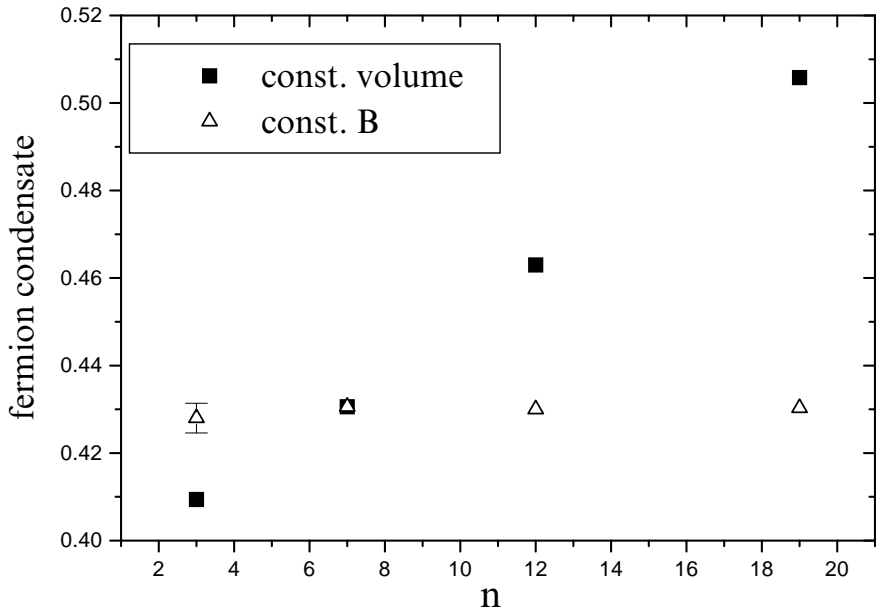,height=10cm,angle=-0}}
\caption{Fermion condensate versus $n$ for $m=0.1$ and $\bt_G =0.30.$ 
The solid squares are the results for a $12^3$ lattice. The triangles
are data for which the volumes are different for each $n.$ 
\label{physb.ps}}
\end{figure}

The question of how to identify the continuum magnetic field strength
$B_{phys}$
requires careful discussion.
In the physical problem we wish to address the
magnetic field is actually the $z$ component of a four-dimensional vector
field,
implying a relation $B=ea\vert\nabla\times{\bf A}^{ext}\vert=e a^2 B_{phys},$
where $e$ is the electronic
charge and is dimensionless. Therefore the combination
$\beta_G^2B$ is dimensionless.
The physical field may
go to infinity letting the lattice spacing $a$ go to zero,
while $B$ is kept constant.

On the lattice the magnetic field is determined by the number $n,$ but 
the allowed values of this number depend on the lattice size $N.$ 
On the contrary, 
the parameter $B$ does not depend much on the details of the discretization, 
so it is natural to choose it as the most natural lattice unit for the 
magnetic field. In figure \ref{physb.ps} we show the fermion condensate 
versus $n.$ The solid squares represent the condensate for a $12^3$ 
lattice for several values of $n.$ We observe that the condensate 
increases substantially with $n,$ since different $n$ yield different 
values for $B.$
The triangles represent data produced by taking each $n$ with an 
appropriate volume, such that $B$ remains almost constant; the resulting 
$(n,N^3)$ pairs turn out to be: $(3,8^3),~(7,12^3),~(12,16^3),~(19,20^3).$ 
It is 
obvious that in this way we get essentially the same condensate for all $n.$ 

An important remark is that the magnetic field should not be 
allowed to grow too
big in lattice units, since then the perturbative expansion of the expansions 
$\exp(i \alpha^{ext}_{x\mu})$ would yield significant $B^2,~B^3,\dots$ 
contributions with the accompanying vertices, in addition to the
desired terms which are linear in $B.$ A trivial estimate of the critical 
field strength is obtained from the demand that the cyclotron radius 
corresponding to a given magnetic field should not be less than (say) two
lattice spacings. This calculation yields $B < \frac{\pi}{8}.$
Of course the above limitations apply strictly only to the case where the
statistical gauge field has been turned off; in the interacting case, 
one does not really know whether there exists a critical magnetic field 
after which discretization effects are important.
With this remark in mind, we depict in the figures of the following sections 
the results for the whole range of the magnetic field, from 0 to $\pi.$

\section{Simulation results}
\label{sec:results}

\subsection{$B=0$}
\label{subs:B=0}

\begin{figure}[h!]
\centerline{
\setlength\epsfxsize{350pt}
\epsfbox{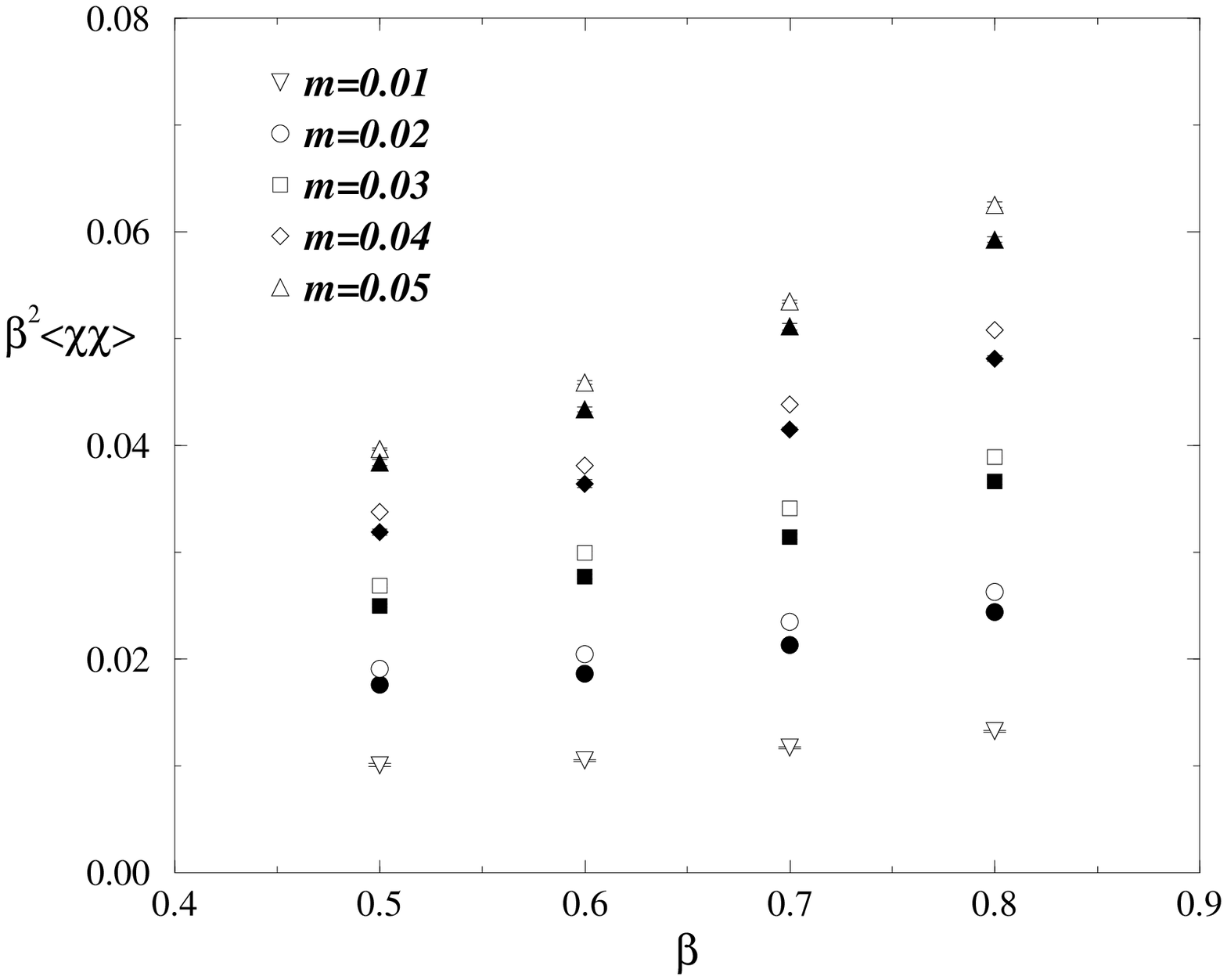}}
\caption{$\beta_G^2\langle\bar\chi\chi\rangle$ vs. $\beta_G$ on 
$12^3$ (filled) and $16^3$ (open) lattices.
\label{fig:simon1}}
\vfill
\centerline{
\setlength\epsfxsize{350pt}
\epsfbox{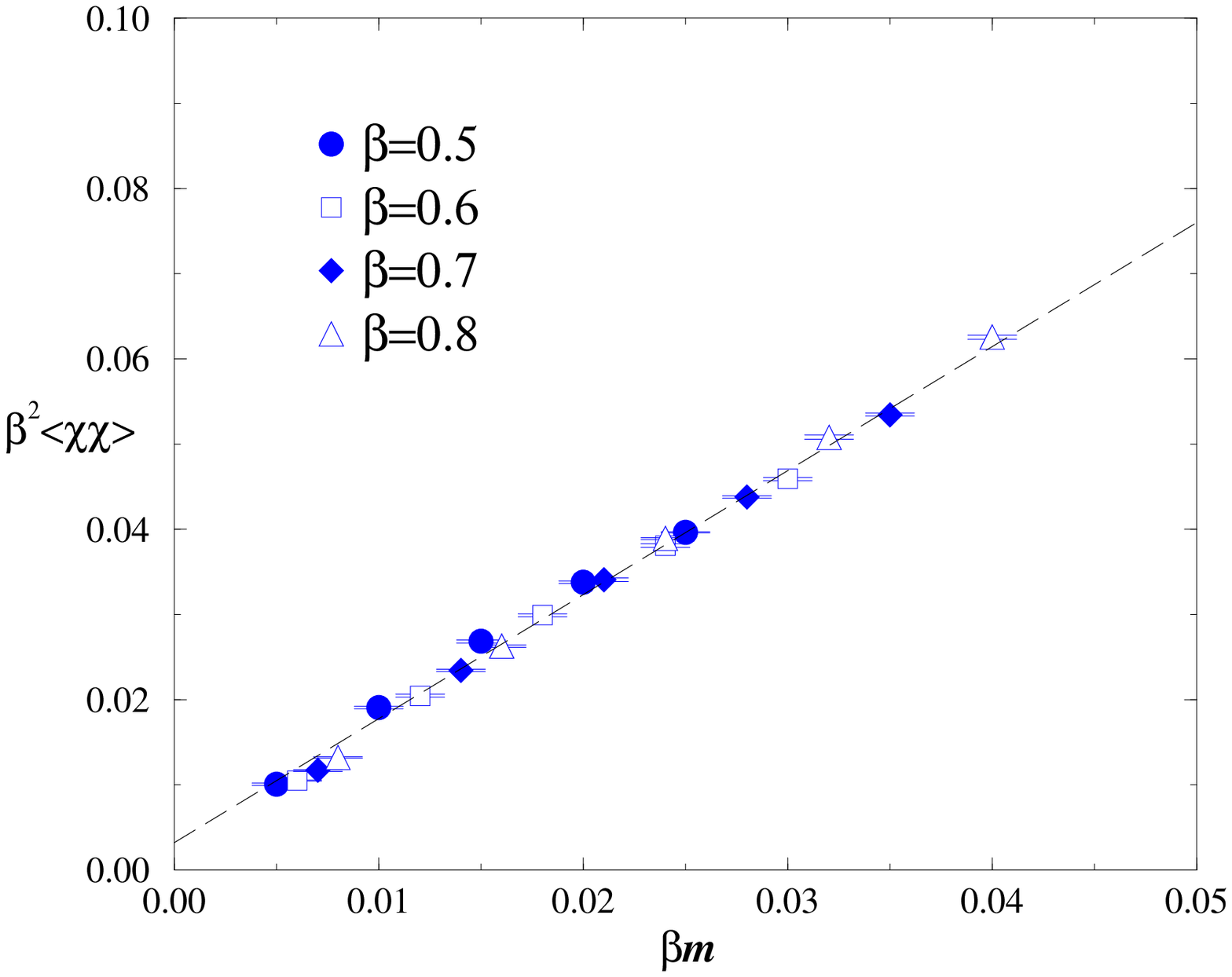}}
\caption{$\beta_G^2\langle\bar\chi\chi\rangle$ vs. $\beta_Gm$
from a $16^3$ lattice.
\label{fig:simon2}}
\end{figure}

It is instructive to begin by studying the case $B=0$. We have performed
simulations on $12^3$ and $16^3$ lattices for a range of gauge couplings 
$\beta_G\in[0.5,0.8]$ and bare fermion masses $m\in[0.01,0.05]$.
For each parameter set we accumulated
$O(2000)$ HMC trajectories of mean length 0.9, with timesteps ranging from 
$d\tau=0.05$ on $12^3$ with $m=0.05$ to $d\tau=0.025$ on $16^3$ with $m=0.01$.
The chiral condensate $\langle\bar\chi\chi\rangle$ was estimated by stochastic
means using 10 noise vectors every two trajectories. Our data is shown in
Fig.~\ref{fig:simon1}. It is clear that finite volume effects are significant;
this can be attributed to the presence of a massless particle, the photon,
in the spectrum of the theory. Large finite volume effects are the main source
of difficulty in numerical studies of QED$_3$ \cite{kogut, azcoiti}. 
We will 
not attempt to extrapolate to the thermodynamic limit, but instead focus on the
$16^3$ results.

Note that in Fig.~\ref{fig:simon1} we have chosen to plot 
the dimensionless combination
$\beta_G^2\langle\bar\chi\chi\rangle$. Since QED$_3$ is an asymptotically-free 
theory (recall $g$ has positive mass dimension), its continuum limit lies
in the limit $\beta_G\to\infty$. To see whether lattice data taken with 
$a\not=0$ are characteristic of the continuum limit, it is helpful to
plot dimensionless quantities against each other; results from different
$\beta_G$ should fall on a universal curve if continuum physics is being probed.
We show a plot of $\beta_G^2\langle\bar\chi\chi\rangle$ 
vs. $\beta_Gm$ in
Fig.~\ref{fig:simon2}. The data do indeed fall 
approximately on a straight line, 
the main departures being at the strongest coupling $\beta_G=0.5$, and
the smallest mass $m=0.01$, which might plausibly be attributed respectively to 
lattice artifacts and finite volume effects. If we exclude these points, 
a straight line fit of the form 
$\beta_G^2\langle\bar\chi\chi\rangle=0.00327(15)+1.454(6)\beta_Gm$ is 
achieved (if the equivalent points from $12^3$ are used, the fitted line 
is $0.00140(16)+1.436\beta_Gm$).

The main result of these fits is that the value of
$\beta_G^2\langle\bar\chi\chi\rangle$ extrapolated to the chiral limit 
is unusually small, but significantly different from zero. 
If the linear form really
does characterise the continuum limit, then this result is robust even on 
our admittedly small volumes.  The implication is that chiral symmetry 
is broken in the the IR limit for $N_f=2$ flavors of fermion in QED$_3$,
and hence $N_{fc}>2$.
This is consistent with the findings of most studies using the gap equation,
but contradicts a recent bound $N_{fc}\leq\frac{3}{2}$
derived by counting massless degrees of freedom 
via the absolute value of the thermodynamic free energy $f$ 
in both IR and UV limits \cite{ACS}, and demanding
\be
f_{IR}\leq f_{UV}.
\ee 
A possible resolution follows from the discussion below eqn.~(\ref{eq:latsym});
the global symmetry of the lattice model is smaller than that of continuum
QED$_3$, resulting in a smaller number of Goldstone bosons in the chirally
broken phase. For $N$ flavors of staggered fermion, there are $8N=4N_f$
fermionic degrees of freedom in the UV limit, and $N^2$ Goldstone boson
degrees of freedom in the IR. Following the argument of \cite{ACS}, we arrive
at the bound $N_c\leq6$ for the lattice model, 
which clearly is not threatened by our results. To substantiate this it will
be necessary to show that the strongly-coupled dynamics of the IR limit is 
governed by the symmetries of the lattice model rather than the continuum one
\cite{thirring}.

\begin{figure}[t]
\centerline{
\setlength\epsfxsize{350pt}
\epsfbox{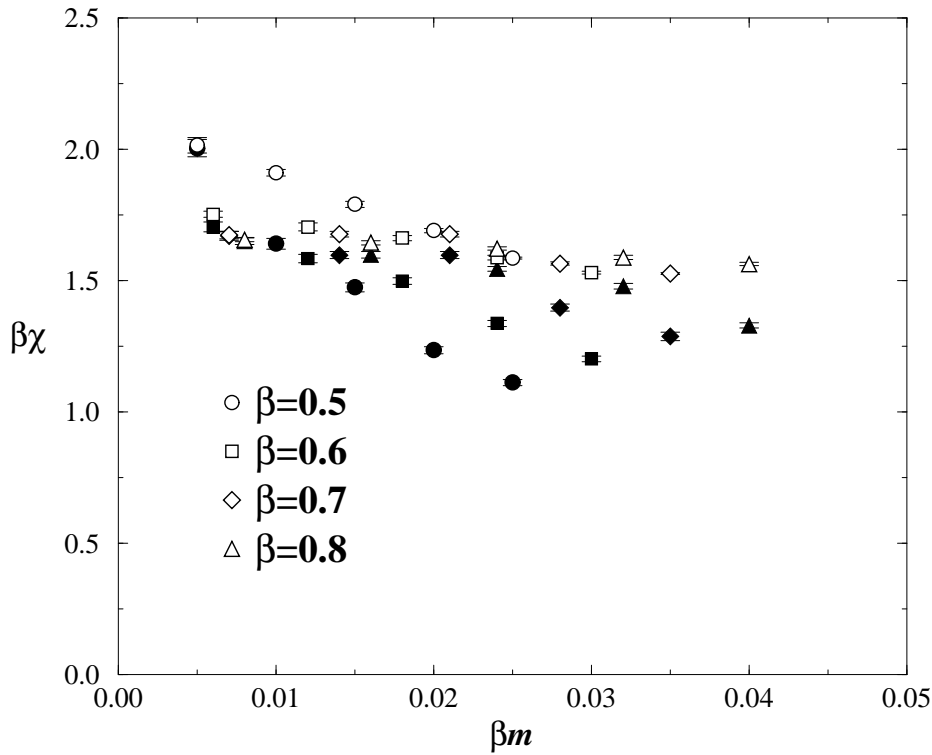}}
\caption{$\beta_G\chi_l$ (filled) and $\beta_G\chi_t$ (open) vs. $\beta_Gm$
from a $16^3$ lattice.
\label{fig:simon3}}
\end{figure}
Next we discuss longitudinal and transverse susceptibilities, defined 
respectively as the 
integrated propagators in scalar and pseudoscalar meson channels:
\be
\chi_l=\sum_x\langle\bar\chi\chi(0)\bar\chi\chi(x)\rangle;\qquad
\chi_t=\sum_x\langle\bar\chi\varepsilon\chi(0)\bar\chi\varepsilon\chi(x)
\rangle.
\label{eq:chis}
\ee
The longitudinal susceptibility $\chi_l$ has contributions from diagrams
with both connected and disconnected fermion lines \cite{thirring}; 
we find that the 
connected contribution is overwhelmingly dominant throughout our parameter
space. The transverse susceptibility is most conveniently calculated via
the axial Ward identity $\chi_t=\langle\bar\chi\chi\rangle/m$.
The results are plotted using dimensionless quantities
in Fig.~\ref{fig:simon3}. We see that the evidence
for universal scaling behaviour in the dimensionless plots is much less
convincing, particularly for $\chi_l$ for $\beta_Gm\geq0.01$. 
At the smallest $m$ values the two susceptibilities
even coincide within errors, suggesting that chiral symmetry is unbroken.
This is probably because the data
from different $\beta_G$ are taken on different physical volumes, and two-point
functions (\ref{eq:chis}) are prone to finite volume effects,
particularly since a Goldstone boson is anticipated in the transverse channel 
as $m\to0$. We can observe a tendency for $\chi_t$ to increase as
this limit is approached. Bearing in mind the linear fit to the data 
of Fig.~\ref{fig:simon2}, however, we might expect to have to go to mass values
as small as $\beta_Gm\approx0.002$ before the divergence in $\chi_t$ due to 
a Goldstone pole becomes dominant. Results in this regime in both 
continuum and thermodynamic limits will be required for universal statements
about the nature of the IR limit to become feasible.

\subsection{$B\not=0$}
\label{b.neq.0}

In this subsection we present the behaviour of fermionic matter 
under the combined action of the external magnetic field and the 
quantum gauge field. 
Before going on with the specific 
features of these results, let us remark that to facilitate comparison 
with the analytic results \cite{lattice} we measured the magnetic field in 
units of its maximal value: thus we used the parameter $b\in[0,1]$ defined 
by $b \equiv B/B_{max} =B/\pi$. 

\begin{figure}[!h]
\centerline{\psfig{figure=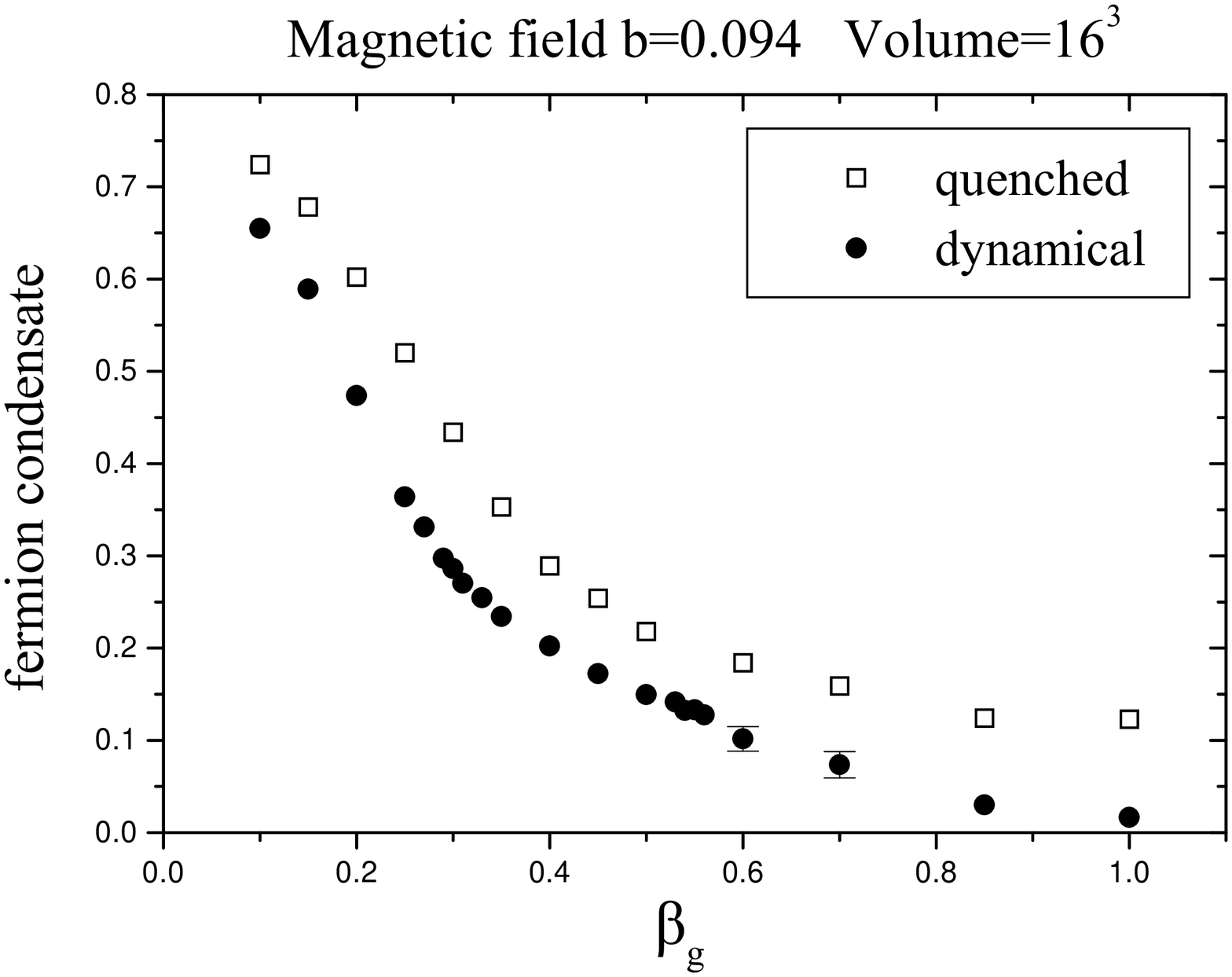,height=10cm,angle=-0}}
\caption{Fermion condensate versus $\beta_G$ for quenched and dynamical
fermions at b=0.094. \label{f2}}
\end{figure}

A comparison of the dynamical
and the quenched results is made in figure \ref{f2}.
A $16^3$ lattice has been used for the calculation,
while the external magnetic field has been set to $b=0.094,$
a typical value.
Measurements have been done at several values of the
bare mass $m$ and what we show in figure \ref{f2} is the 
extrapolation to the $m \to 0$
limit. The method of extrapolation will be explained in more detail in the
discussion of figure \ref{bg033fit.ps} below. The volume
dependence is not very big in the quenched case \cite{fkmm}. 
We observe that the values for the condensate
in the quenched case are rather large for the relatively large value
$\beta_G=1.0$ for the gauge coupling. In principle the condensate should
vanish for $\beta_G \rightarrow \infty,$ since the system moves to the
``free" case, and the lattice volume is fixed. Presumably, even the
quenched condensate will vanish, but the figure suggests that 
this will require a much bigger value of $\beta_G.$ 

The fact that the full condensate 
for dynamical fermions is smaller than the corresponding quantity for 
quenched fermions has presumably to do with 
Pauli repulsion, which has clearly an effect in the dynamical fermion case.

\begin{figure}
\centerline{\psfig{figure=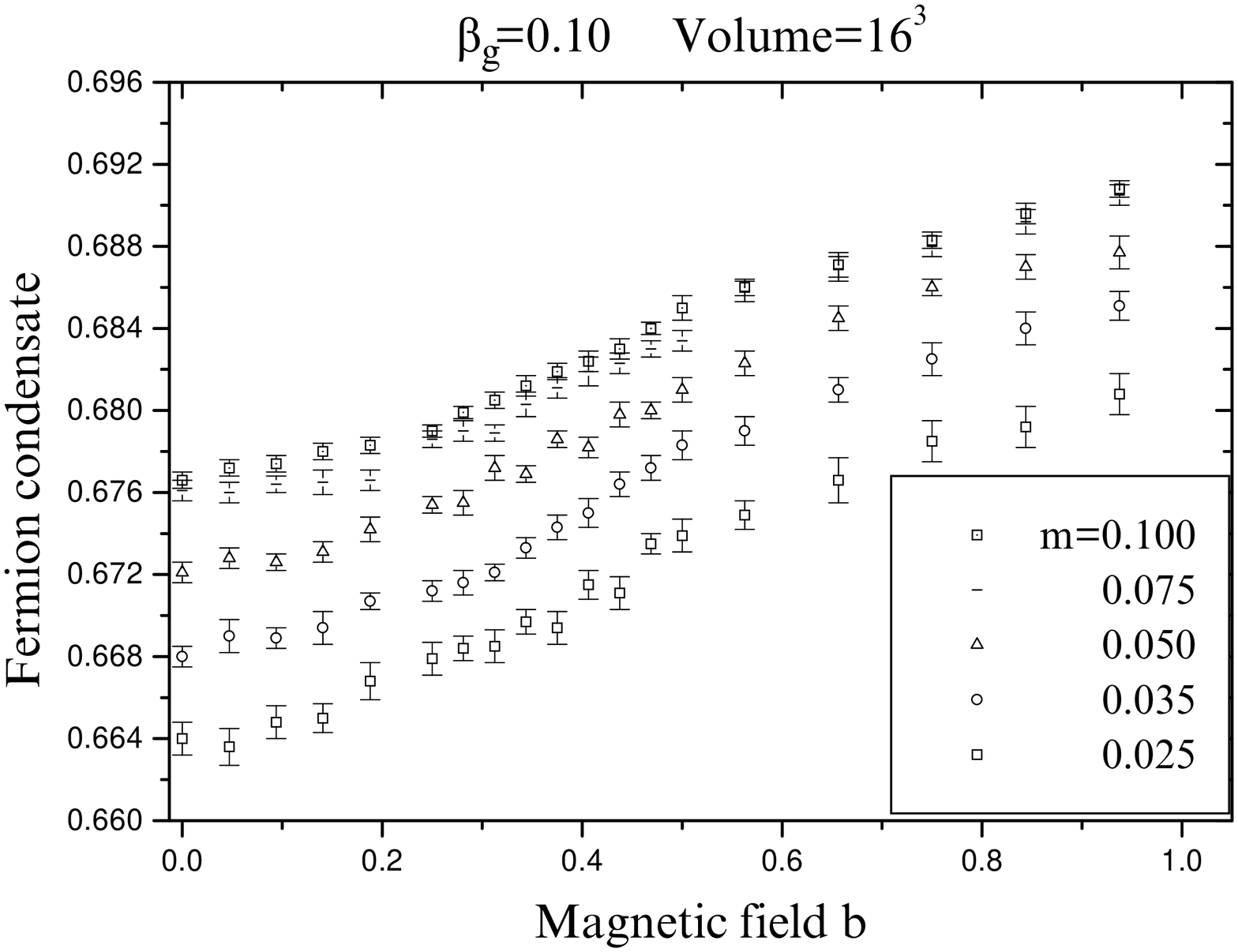,height=10cm,angle=-0}}
\caption{Fermion condensate at strong coupling versus the magnetic field 
for various bare masses. \label{bg0.10m1.ps}}
\vfill
\centerline{\psfig{figure=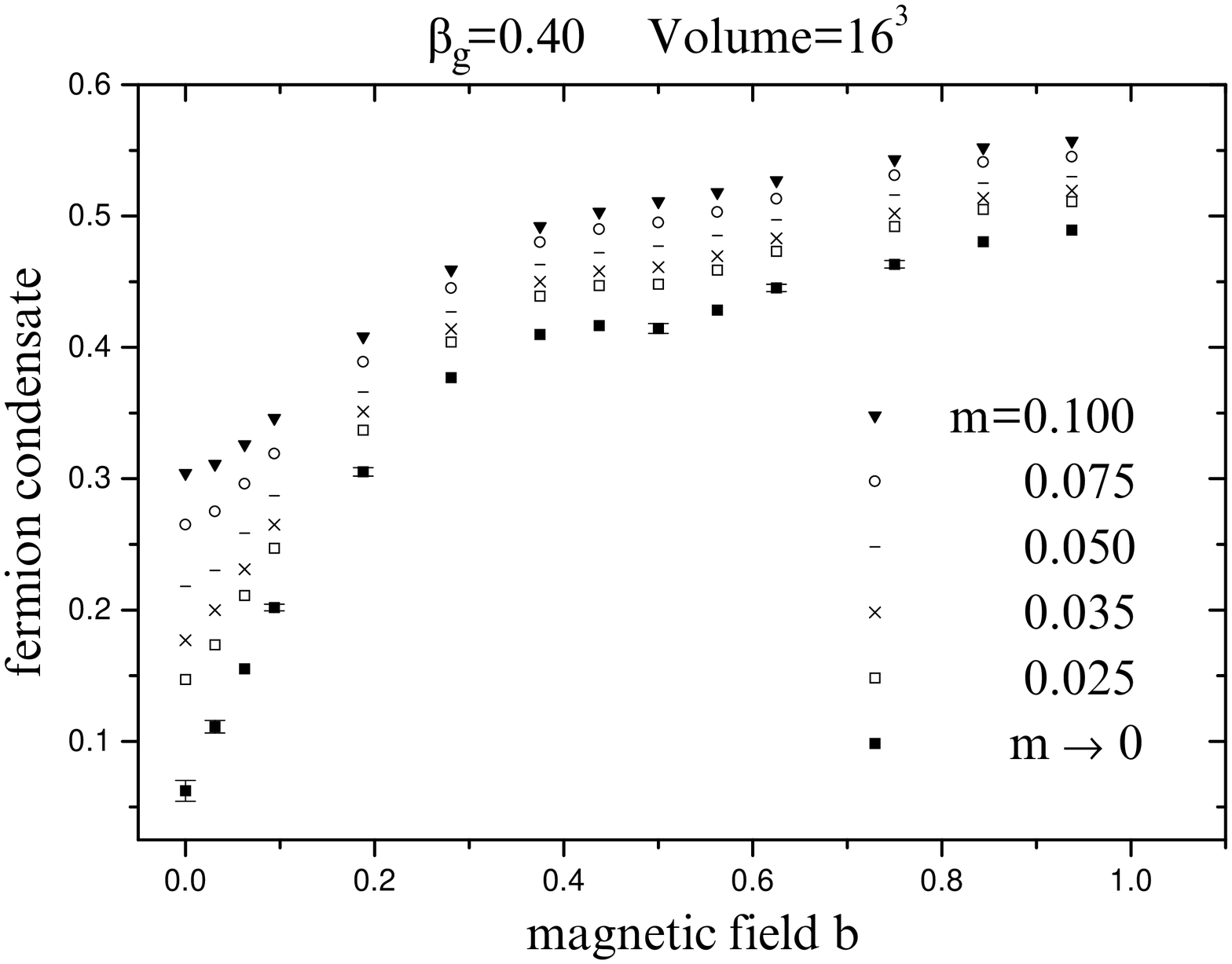,height=10cm,angle=-0}}
\caption{Mass dependence for intermediate $\beta_G.$ 
\label{bg0.40_b.ps}}
\end{figure}

The next set consists of measurements of the fermion condensate 
versus the magnetic field for a $16^3$ lattice at a fixed, strong coupling 
for the statistical gauge field $(\beta_G = 0.10)$, for five  
values of the bare mass (figure \ref{bg0.10m1.ps}). 
For all five masses 
the plot consists of
two parts with different behaviour. For relatively small $b$ 
we find a dependence of the condensate on the external
magnetic field which is nearly linear. 
For relatively big magnetic fields we find points that
have a qualitatively different behaviour.  
The two regions merge smoothly. The ``linear" part of the graph 
appears to extend in a wider interval of $b$ for decreasing mass.  
It should be stressed that the 
mass dependence of the condensate is in sharp disagreement with the
corresponding result in \cite{fkmm}, where quenched fermions 
have been used. The condensate {\em increases}
with the bare mass for all values of the magnetic field strength, 
whereas in the quenched case it was {\em decreasing.} The $m \to 0$ 
extrapolation of the results lies probably under the lowest curve, 
however at this stage our statistics are not good enough to get reliable 
numbers for the limiting values of the condensate.

It is interesting to know what happens as one moves towards weaker couplings.
Based on the results of the quenched case \cite{fkmm} we expect that the
almost linear part referred to previously would be restricted to
very small magnetic fields and eventually disappear. 
In figure \ref{bg0.40_b.ps}
we show the results for the coupling $\beta_G=0.40.$ The linear part 
has indeed been restricted to the region $b<0.2.$ 
Moreover, we show the bare mass dependence along
with the $m \to 0$ limit of the condensate versus $b.$ It is 
interesting to note that the data for each particular mass give a 
monotonic dependence, while the $m \to 0$ limit 
starts having a local minimum at $b\simeq0.5$, 
which is characteristic of the free
case, ie. with the statistical gauge field turned off, to which we now turn. 
\begin{figure}[!h]
\centerline{\psfig{figure=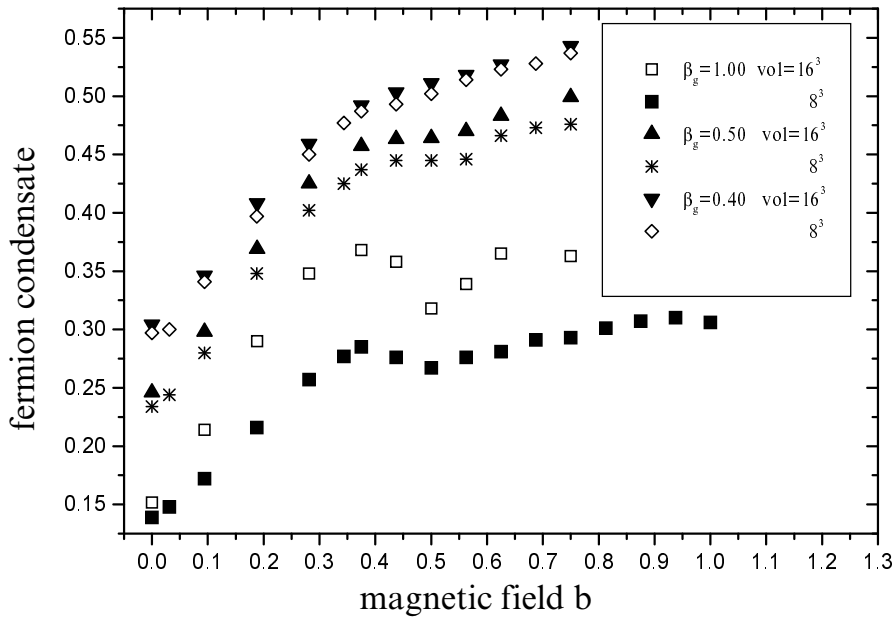,height=10cm,angle=-0}}
\caption{Approach to the free case in the limit of large
$\beta_G$ and a fixed fermion mass $m=0.100.$ 
\label{m0100_fin_siz.ps}}
\end{figure}
The free case 
was studied in \cite{lattice}, where it was found 
that for big enough $b$ the condensate stopped being monotonic and started 
showing local maxima and minima. 
It developed the 
first local minimum at $b=0.5$ and then it had a
succession of maxima and minima, up to $b=1.$ Moreover, there was 
a spectacular volume dependence. One expects, of course, that this 
free case will be reached for big enough $\beta_G.$ 
\begin{figure}[!h]
\centerline{\psfig{figure=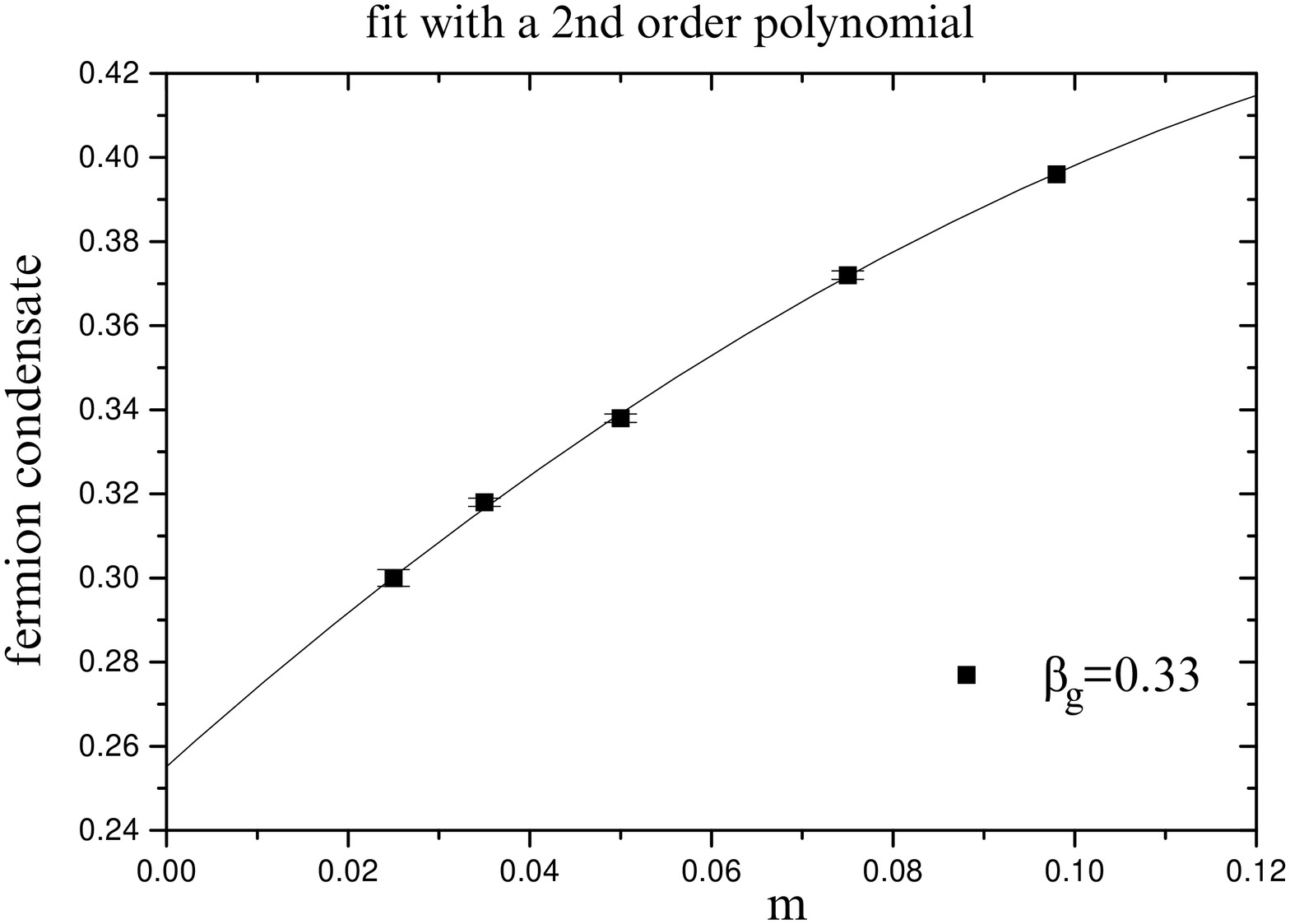,height=10cm,angle=-0}}
\caption{Extrapolation to the zero mass limit for $\beta_G = 0.33$
and b=0.094. \label{bg033fit.ps}}
\vfill
\centerline{\psfig{figure=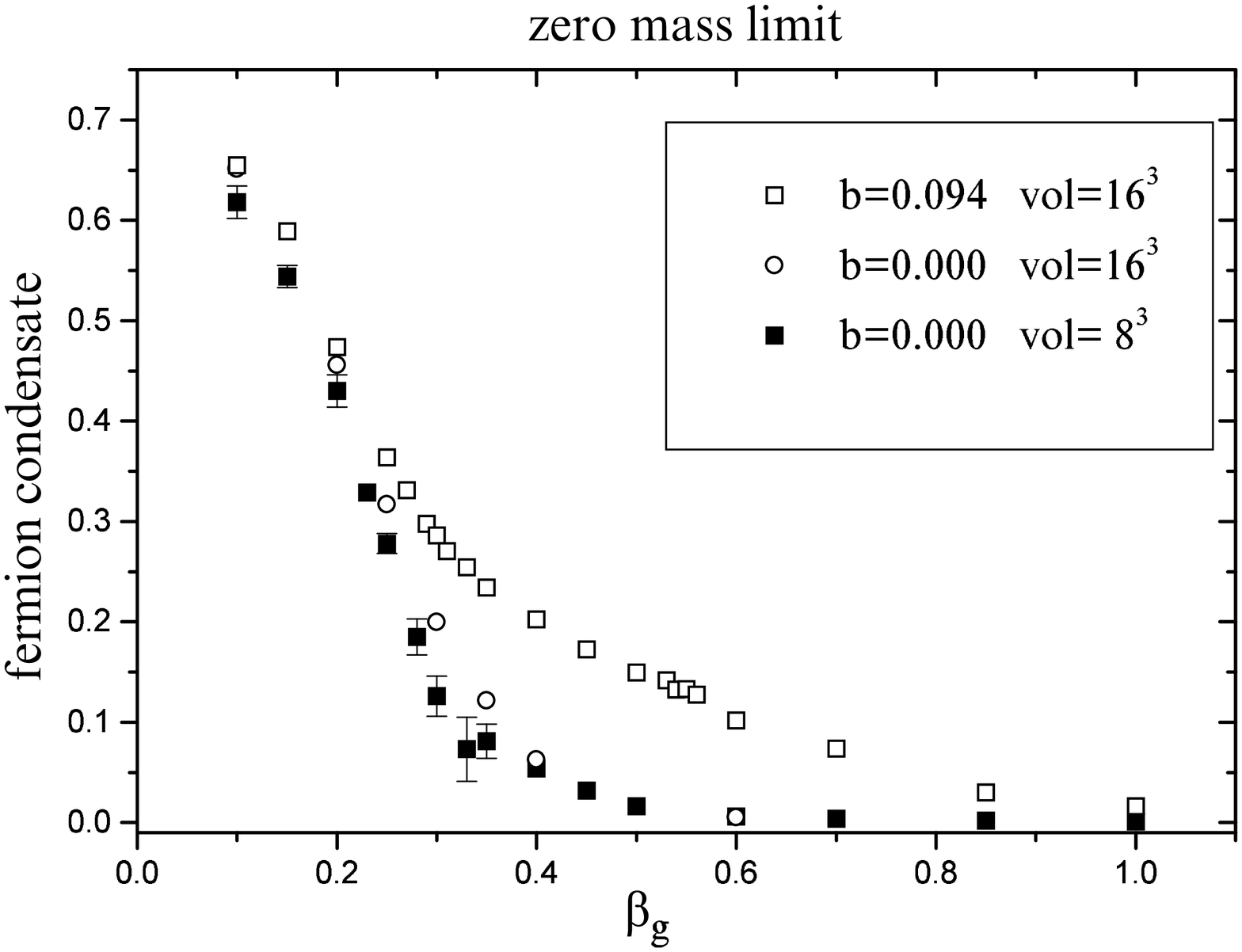,height=10cm,angle=-0}}
\caption{$\beta_G$--dependence of the condensate for two
values of the magnetic field: $b=0$ and $b=0.094.$ \label{b0b12dyn.ps}}
\end{figure}
We have seen this kind of approach to the free model 
in the quenched fermion case \cite{fkmm}.
In figure \ref{m0100_fin_siz.ps} we show the results 
for $\beta_G=0.40$, 0.50 and 1.00 for two volumes, 
$8^3$ and $16^3.$ The coupling $\beta_G=0.50$ is the point where
the curve shows the first sign of non-monotonous behaviour. 
At $\beta_G=1.00$ the successive maxima and minima are
clear. These features are shared with the quenched case. 
However, in the quenched calculation the volume dependence was not visible, 
while in this case, one can see the volume dependence at all values 
of the parameters. In the intermediate coupling regime 
($\beta_G=0.40$ and $\beta_G=0.50$) the volume dependence,
although clear,  is rather small; at $\beta_G=1.00$ the volume dependence 
is comparable with that of the free case. In view of this behaviour we
may be confident that, at this large $\beta_G,$ the limit of switching the
gauge field off can be reached if one uses dynamical fermions, in 
contradistinction with the quenched calculation \cite{fkmm}. 
One should remark that in the free 
case the role of the bare mass is very important, since it is eventually
the only source of mass generation. This is at the root of the large
volume dependence showing up in the free case: at fixed volume  
the condensate goes over to zero for vanishing bare mass. In the 
quenched interacting model, though, the interaction with the 
gauge field generates a 
dynamical mass, independently from the value of the bare mass. 
The volume dependence is thus small, permitting a 
smooth transition to the thermodynamic, as well as to the massless, limit 
\cite{HK}.
In the present case with the dynamical fermions, the dynamically generated 
mass is suppressed, presumably due to the Pauli repulsion which is in full
operation here, so the value of $\beta_G$ for which we have the transition 
to the free case is rather small compared to the quenched
calculation.

Now we are going to explain in some detail the procedure we followed
to obtain the $m \to 0$ limit. 
The simulations have been done at non-zero values of the bare mass; the 
massless limit has been taken by extrapolating the results for several 
bare masses to the limit $m \to 0.$ Figure \ref{bg033fit.ps} 
shows the process of this extrapolation for $\beta_G=0.33$ and the 
external magnetic field $b$ set equal to 0.094. The general behaviour 
is similar to that of the quenched case, in the sense that
in the strong coupling region this curve is an almost straight line with 
a small slope and in the weak gauge coupling a second order polynomial 
proves necessary to perform the fit. 
At strong coupling the mass dependence is not
very big, because it is the strong gauge coupling which dominates
in the formation of the condensate. The error bars appear to be smaller 
here as compared to the gauge couplings of similar magnitude 
in the quenched case \cite{fkmm}. 

\begin{figure}[h!]
\centerline{\psfig{figure=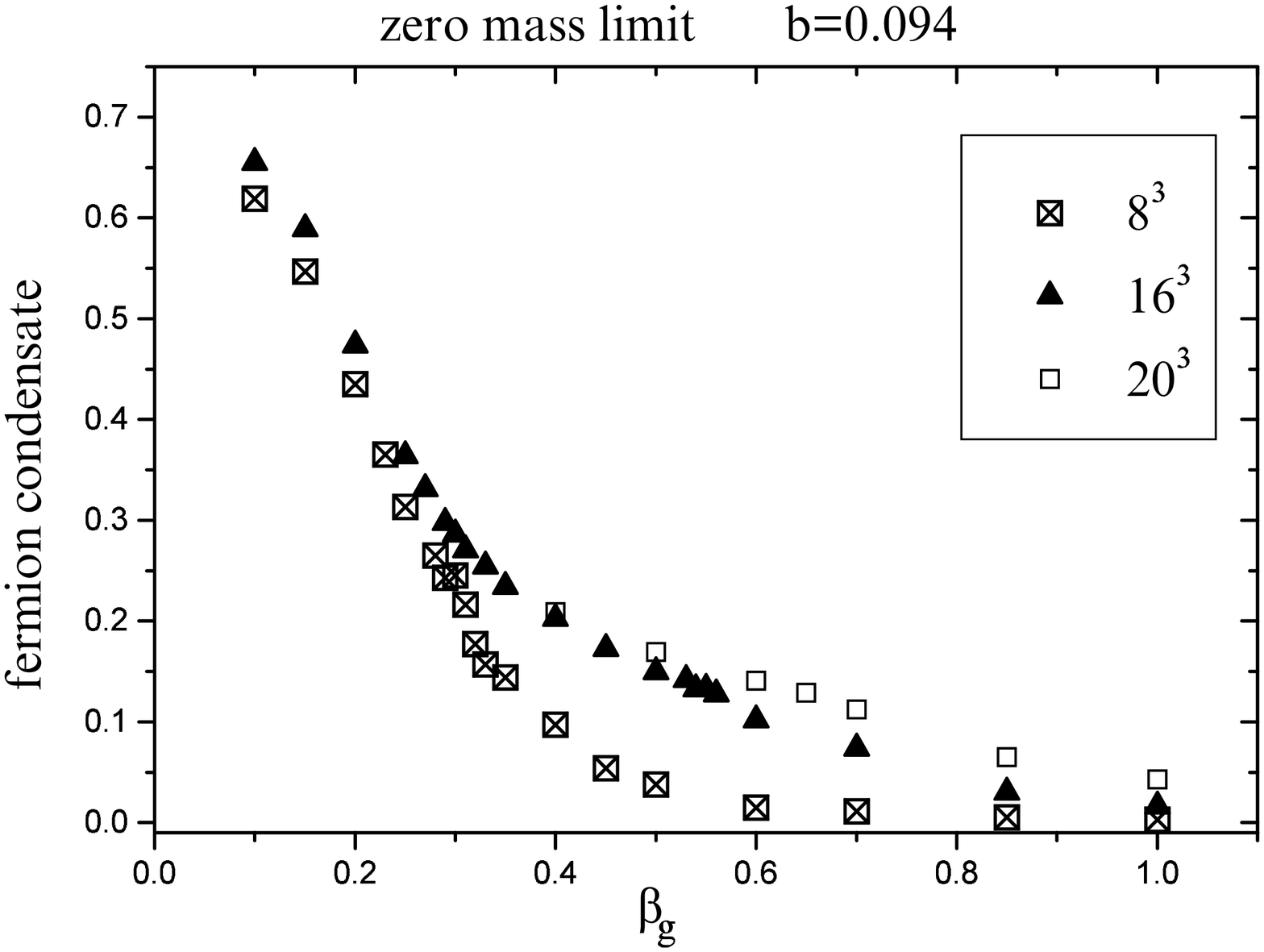,height=10cm,angle=-0}}
\caption{Volume dependence and finite size effects for dynamical 
fermions in the zero mass limit. \label{b12_fin_siz.ps}}
\vfill
\centerline{\psfig{figure=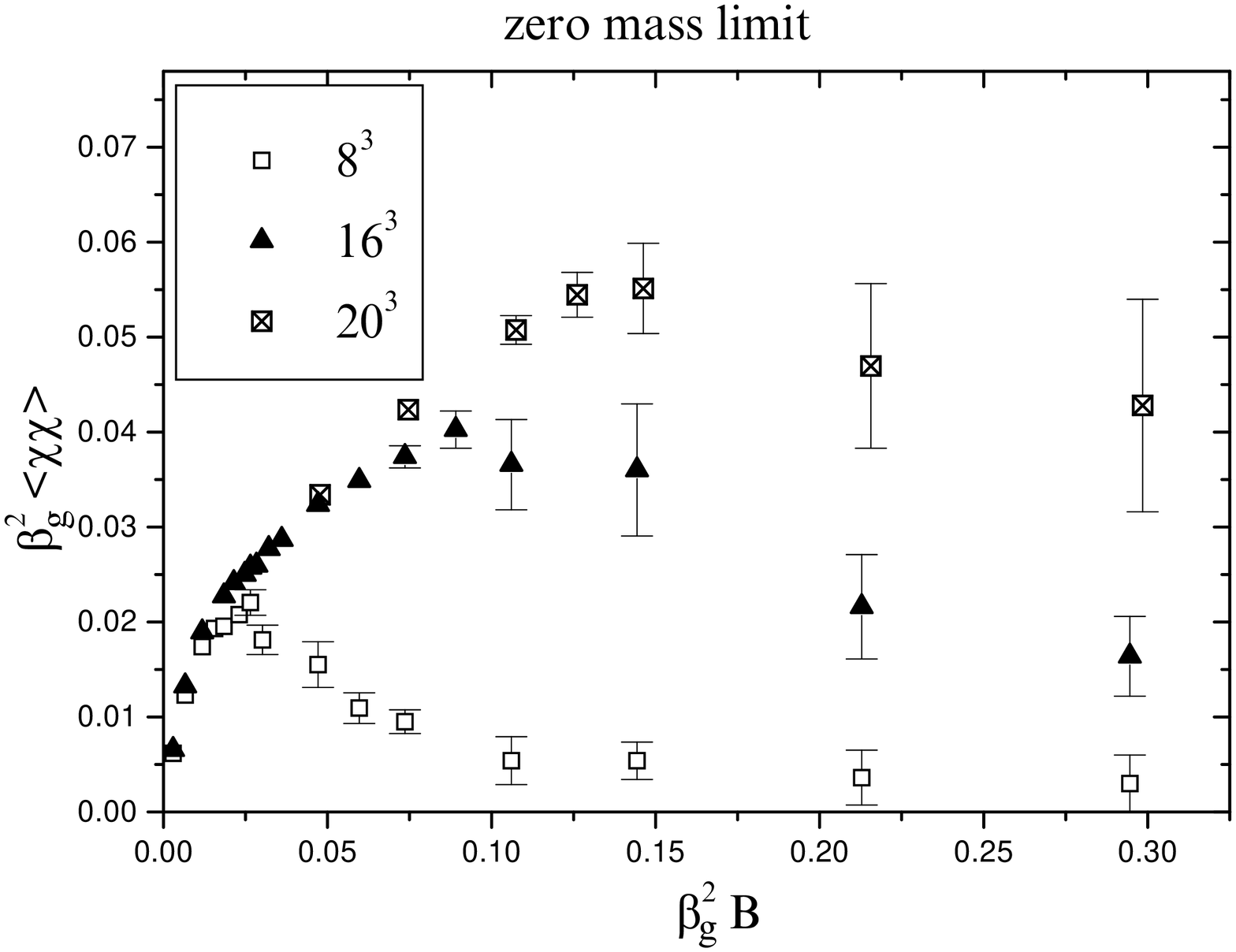,height=10cm,angle=-0}}
\caption{Finite size effects for dynamical 
fermions in the zero mass limit using dimensionless variables. 
\label{dimensionless.eps}}
\end{figure}

Figure \ref{b0b12dyn.ps} contains the zero mass limit of the 
condensate (obtained through the procedure illustrated in figure 
\ref{bg033fit.ps}) versus $\beta_G$ 
for $b=0$ and $b=0.094.$ In the former case the bulk of the results 
come from $8^3$ lattice, however we also present some results 
from $16^3,$ just to get some feeling about the volume dependence. 
We observe that in the strong coupling region the $b$-dependence of 
the condensate is rather weak; on the 
contrary, at weak coupling, the external field is the main 
generator of the condensate, and we find an increasingly big 
$b$-dependence as we move to large $\beta_G.$ In particular for zero 
external field the result is very small for $\beta_G > 0.5,$ while the 
external field at these values of the gauge coupling still generates 
a relatively big condensate. 

The volume dependence of the condensate at $b=0.094$ for a wide range
of $\beta_G$ is shown in figure \ref{b12_fin_siz.ps}. 
For the smallest volume $8^3$ the data exhibit a rather big cusp
at $\beta_G \simeq 0.30.$ For the two bigger volumes $16^3$ and $20^3$
the slopes are discontinuous at $\beta_G \simeq 0.53$
and $\beta_G \simeq 0.60$ respectively, but the discontinuity is smaller.
It appears to be a finite size 
effect, possibly a shadow of the deconfining transition for each 
volume at weak coupling. The dynamically generated
mass is smaller for the model with dynamical fermions as compared to
the quenched model for the same value of the gauge coupling; so the finite 
size effects become more manifest.

In figure \ref{dimensionless.eps} we rescale the axes of figure
\ref{b12_fin_siz.ps}, such that we get dimensionless units. In
particular, we use $\bt_G^2 B \to e B_{phys}/g^4$ 
for the $x$ axis and $\bt_G^2 \langle{\bar \chi} \chi \rangle
\to \langle{\bar \chi} \chi\rangle\vert_{phys}/g^4$ 
for the $y$ axis. What was a cusp in figure 
\ref{b12_fin_siz.ps} has been transformed into a {\it maximum} 
in dimensionless units. The condensate starts increasing with the
external magnetic field, now that it is properly rescaled, but finally
the finite size effects take over and the condensate starts decreasing.
The value of the magnetic field where this happens increases with the 
lattice volume and presumably tends to infinity in the thermodynamic 
limit. 

\section{Summary}

After reviewing and contrasting the formulation and 
global symmetries of QED$_3$ in continuum field theory and on the lattice, we
have presented numerical results from a Monte Carlo simulation study using
dynamical fermions. 
For $B=0$ our calculations used the largest system sizes studied to date to our
knowledge; for $B\not=0$ this is the first such study. Our principal results
are as follows:

\begin{itemize}

\item For $B=0$ we have evidence, on the assumption of a linear scaling
relation, for spontaneously broken chiral
symmetry in the chiral and thermodynamic limits for $N_f=2$ flavors of
four-component fermion. The condensate $\langle\bar\chi\chi\rangle$
has an unusually small value, 
$O(10^{-3})$ in natural units, which perhaps explains why this
issue has proved a ``difficult'' problem in non-perturbative field theory.
Our results should inform future studies on the system sizes needed in
order for universal statements about the continuum theory to be possible.

\item  We have studied the response of the condensate to an external magnetic
field in the regimes of strong, intermediate and weak U(1)$_S$ coupling. In all
cases $\langle\bar\chi\chi\rangle$ is smaller than that found in comparable
quenched simulations. As in
the quenched case, for strong and intermediate coupling regimes the condensate
increases monotonically with $B$, but the behaviour becomes non-monotonic 
at weaker coupling. 
In the weak coupling limit contact with the behaviour of free fermions 
including volume
dependence is recovered, in contrast with the quenched model.

\item We have used our results in the chiral limit to make a plot of condensate
against external field in dimensionless units (Fig. \ref{dimensionless.eps}), 
which enables a significant finite volume effect to be identified.
The condensate increases as a smooth monotonic function of $\beta_G^2B$ 
with negative
curvature. Future studies with different values of the lattice magnetic field
parameter $b$ will help to establish whether this behaviour is universal, and
hence characterises the continuum limit.
We note that in the Schwinger-Dyson approach \cite{jalex} 
the regime of
small magnetic fields could not be efficiently treated, while this
is feasible with the Monte Carlo approach.

\end{itemize}

\section*{Acknowledgements}

This work has been done within the TMR project ``Finite temperature 
phase transitions in particle physics", EU contract number: FMRX-CT97-0122.
The authors would like to acknowledge financial support from the
TMR project, and S.J.H. also from the Leverhulme Trust. 
G.K. would like to thank P. Dimopoulos for help with the graphics, and S.J.H.
Ian Aitchison for helpful discussions.

\end{document}